# High quality factor gigahertz frequencies in nanomechanical diamond resonators

Alexei Gaidarzhy
*Department of Aerospace and Mechanical Engineering, Boston University, 110 Cummington Street, Boston, Massachusetts 02215*
Matthias Imboden and Pritiraj Mohanty [a)]
*Department of Physics, Boston University, 590 Commonwealth, Boston, Massachusetts 02215*
Janet Rankin and Brian W. Sheldon
*Division of Engineering, Brown University, Providence, Rhode Island 02912*

**We report actuation and detection of gigahertz-range resonance frequencies in nano-crystalline diamond mechanical resonators. High order transverse vibration modes are measured in coupled-beam resonators exhibiting frequencies up to 1.441 GHz. The cantilever-array design of the resonators translates the gigahertz-range resonant motion of micron-long cantilever elements to the displacement of the central supporting structure. Use of nano-crystalline diamond further increases the frequency compared to single crystal silicon by a factor of three. High clamping losses usually associated with micron-sized straight beams are suppressed in the periodic geometry of our resonators, allowing for high quality factors exceeding 20,000 above 500 MHz.**

The realization of gigahertz-range mechanical resonance frequencies and high quality factors is of interest for a wide range of technical and fundamental research. Applications range from frequency selective oscillators and passive filters for microwave communications[1], high speed mechanical memory elements[2], to ultrasensitive mass or charge detectors[3]. From a fundamental physics perspective, demonstration of quantum corrections in the motion of nanomechanical resonators could be achieved using ultra-high frequency devices at millikelvin temperatures[4]. Here, we demonstrate high quality factor gigahertz range nanomechanical resonators by combining the high Young's modulus of nanocrystalline diamond[5] and recently developed coupled-element resonator architecture[6].

Previously we have reported that the addition of arrays of short cantilever elements to a much longer doubly clamped beam results in an enhancement of the net response of certain high order resonant modes[6]. These collective modes arise from in-phase resonant motion of the array elements. The cooperative action of these elements drives the central beam, generating large net displacements and signals by effectively reducing the motional stiffness on resonance. The resulting performance of the coupled-beam resonator offers significant advantages when compared with straight beam designs in the gigahertz frequency range.

In addition to the resonator geometry, the use of diamond as the structural material has obvious advantages over other commonly used materials in NEMS and MEMS fabrication, such as single crystal silicon and silicon carbide. In our submicron-thin nano-crystalline diamond films we measure values for the Young's modulus ($E$) on the order of 900 GPa, which together with density ($\rho$) translates into a factor of nearly three increase in the material sound velocity $v_s = \sqrt{\dfrac{E}{\rho}}$ and frequency compared to silicon[5]. Particularly relevant to low temperature studies is the high thermal conductance[7] of diamond, which allows for efficient thermalization of nanostructures.

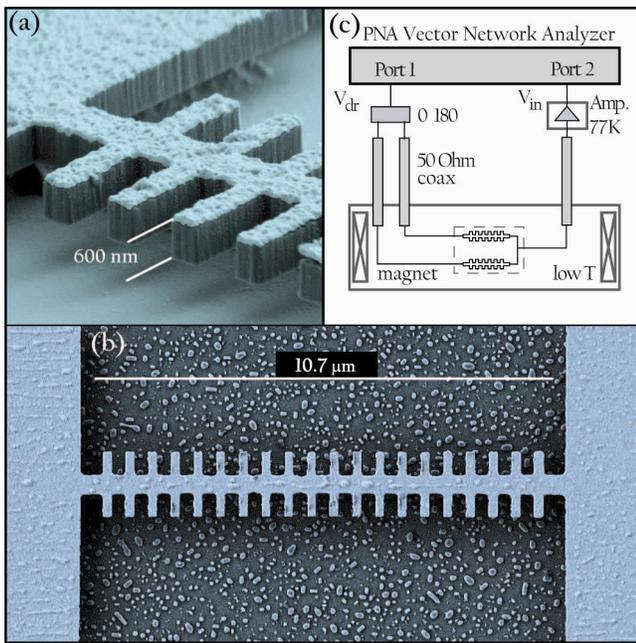

**Figure 1**. (color online) Measurement electronics setup and SEM pictures of diamond antenna structure. **(a)** Close up view of antenna structure. The gold electrode used for actuation and detection is the 50 nm thin light colored layer on top of the diamond. The diamond surface roughness is of the same magnitude as the electrode thickness. **(b)** Top view of entire antenna structure. Samples with two central beam lengths of 10.7 and 21.4 microns are measured, each containing two rows 20 perpendicular paddles. **(c)** PNA vector network analyzer allows for actuation and detection of both amplitude and phase. The dilution refrigerator enables temperatures as low as 40mK. The double bridge technique is implemented using the phase splitter. The output from the device under test is amplified with a low noise cryogenic 24 dB pre-amplifier.

The nano-crystalline diamond (NCD) films used to fabricate the resonators are grown by microwave plasma CVD (AsTex HPM/M system) with a gas mixture of 1% $CH_4$, 5% $H_2$, and 94% Ar. Standard e-beam lithography and surface nanomachining (RIE and buffered HF procedures) is used to fabricate the structures. The growth conditions for the diamond films varied from sample to sample due to fluctuation in seeding density and location in the growth chamber. More detailed descriptions of the NCD produced and fabrication process used are provided elsewhere[5,8]. The SEM micrographs (a) and (b) in Figure 1 display diamond resonators fabricated by this method. Two device scales are studied; the larger structure contains a central beam 21.1 μm long and 400 nm wide. The device contains 20 cantilevers with width 300 nm and length 500 nm, perpendicular to the central beam. The smaller structure is scaled by a factor of 0.5 in all dimensions with the exception of the beam thickness.

The thickness of NCD diamond films used in this work varies from 400 to 600 nm, with estimated post processing RMS surface roughness of 50 nm. Device fabrication induces additional roughness, since the as-grown films have RMS roughness values of ~10-15 nm (on the order of or slightly larger than the grain size). The roughness introduces an effective average thickness over the entire beam that may vary from the local thickness of an individual paddle. In the high frequency collective modes reported here, the film roughness has the effect of increasing the spread in the natural frequencies of the otherwise identically designed cantilevers, as the frequency is proportional to the beam thickness to first order.

Residual stresses in the NCD films are also likely to have an impact on their performance. The thermal expansion mismatch between the film and the Si substrate produces roughly -0.4 GPa of compressive stress, as the sample is cooled from the deposition temperature. The growth process produces intrinsic tensile stress which counteracts some of this thermal stress[8], such that films grown under the conditions used here have a total stress below roughly -0.1 GPa. The growth process also produces stress gradients, with increasing tensile stress as the film thickness increases[8]. Because the films used here are relatively thin, these gradients are relatively modest (< ~0.1 GPa difference across the thickness of the film).

In NCD, it is widely believed that $sp^2$ bonded carbon at grain boundaries influences a number of properties[9,10,11,12]. For example, a decreasing elastic modulus with increasing $sp^2$ bonding is consistent with our recent work with these materials[5]. The length scale associated with these modulus variations is on the order of the grain size (< 10 nm), thus it is reasonable to describe the material with a single average modulus value for length scales over a micron. It is not yet clear whether these inhomogeneities will have a significant impact on the losses in our high frequency resonators.

| Sample # | $f$ [MHz] | $Q_{max}$ | $fQ_{max}$ [Hz] |
|---|---|---|---|
| 1 | 7.805 | 3388 | $2.64\times10^{10}$ |
|   | 434.3 | 21500 | $9.34\times10^{12}$ |
| 2 | 7.801 | 27315 | $2.13\times10^{11}$ |
|   | 630.6 | 23200 | $1.46\times10^{13}$ |
| 3 | 37.2 | 13500 | $5.02\times10^{11}$ |
|   | 836.9 | 1075 | $5.20\times10^{12}$ |
| 4 | 20.67 | 7500 | $1.55\times10^{11}$ |
|   | 1441.8 | 8660 | $1.25\times10^{13}$ |

**Table I.** Mode frequencies $f$, with corresponding quality factors $Q$. On every reported sample we measure two prominent modes, the fundamental low MHz-range mode and a collective high frequency mode. The quality factors of the collective modes are very high compared to predicted quality factors of simple beams at similar frequencies. The product of the frequency and quality factor ($fQ_{max}$) rivals the highest measured to date (see Ref. 1).

The antenna structure consists of a long central beam and adjacent paddles that are an order of magnitude smaller. The size difference of these frequency-determining elements makes them sensitive to variations in stresses and stiffness on different length scales. The central beam reflects the average global film properties on the scale of 10 micrometers, whereas the paddles are sensitive to the mechanical properties on the submicron scale. Correspondingly, a nearly 50% variation of the measured collective mode frequencies is observed compared to a finite element simulated structure geometry. The observed fundamental resonances are more accurately predicted. For some samples, crystal defects and surface roughness may not allow for a resonant in-phase actuation of all paddles, and hence the collective mode is not detectable.

The resonators are actuated and the response recorded using the magnetomotive technique. We place the samples under vacuum (< $10^{-6}$ Torr) in a magnetic field, at temperatures ranging from 40 mK to 1.1 K, significantly below the Debye peak of diamond[5]. A low temperature pre-amplifier (77K) is used to amplify the readout signal. The oscillators are actuated in the out of-plane mode (the displacement perpendicular to the substrate), and sensitivity is improved using the parallel bridge technique[5] (measurement setup is illustrated in Figure 1 (c)). The data reported here are all taken in the weak driving regime where the mode response is harmonic. The effective spring constant is extracted from the linear force versus displacement relation measured on resonance (see Figure 2 (b)).

According to finite element modeling, the net displacement of each resonator is greatest in its fundamental resonance mode (typically at low megahertz frequencies), while the response of the next higher harmonics decreases rapidly and quickly falls below the noise floor. Above a few hundred megahertz the enhanced response of the first and subsequent collective modes again become prominent and readily detectable due to the coupled-element dynamics of the resonator[6]. The highest frequency mode for these resonators is measured at 1.44 GHz, see plots in Fig. 2. From all the observed collective resonances this result comes closest to finite element simulations, where the idealized structure is expected to show a collective response at 1.32 GHz. Finite element simulation predicts the next group of collective modes at 4.5 GHz, this lies beyond the current range of the measurement instrumentation. We expect future measurements at higher frequencies to reveal the next group of collective resonances.

Exceptionally high quality factors of 21500 to 23200 for the 430 MHz and 630 MHz collective modes respectively are measured. The smaller resonators, with collective mode resonances at 0.8 and 1.4 GHz, also show unusually high quality factors of 1075 and 8660 respectively (see Table I). An important parameter of MEMS and NEMS resonators for practical applications is the product of the frequency and quality factor (*fQ*). In our devices, values of *fQ* are measured up to $1.46\times10^{13}$ Hz for a collective mode, two orders of magnitude over the *fQ* of the fundamental mode. Our maximum *fQ* is similar to the highest value reported to date for nanomechanical resonators[1].

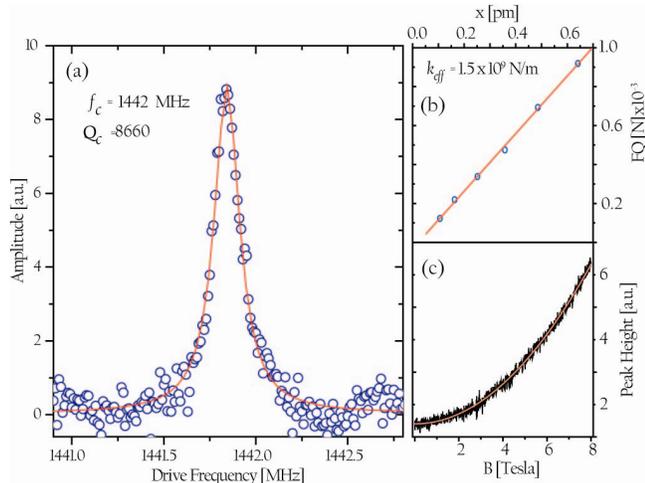

**Figure 2.** (color online) **(a)** Amplitude response of the 1.441 GHz collective mode, including Lorentzian fit (solid red line), from which the quality factor of 4380 is extracted. **(b)** Force vs. displacement plot, obtained from multiple sweeps at varying power. The effective spring constant is determined from the linear fit. **(c)** Quadratic magnetic field dependence of the peak amplitude, including a numerical fit.

The collective mode quality factors contrast with the picture of rapidly increasing energy dissipation generally observed in straight beam resonators with equivalent resonance frequencies[5]. The vanishing of the quality factors at high frequencies has been attributed in part to the increased loss of acoustic energy into the supports of the structure[13]. The high Q-factors observed in our modes signal a possible suppression of clamping loss due to the impedance mismatch along the resonator central beam resulting from the cantilever arrays. The analysis of acoustic wave propagation in periodic structures[14] qualitatively agrees with our findings. The dissipation processes for nanomechanical oscillators at subkelvin temperatures are a topic of great interest and research. Mechanisms other than clamping losses, such as surface defects and tunneling in two level systems are known to be of importance. More data, on a larger number of antenna resonators, must be collected for a comprehensive dissipation analysis, and is the subject of future work.

Our high frequency collective modes, with initial quality factors as high as 21400, demonstrate significant peak broadening (by 200%) and frequency down-shift (by less than one percent) after prolonged high-power actuation and nonlinear measurements (not reported here). Dissipation is also sensitive to experimental conditions such as temperature and magnetic field strength[5,15], which can change the dissipation typically by a factor of 3-4.

Here we have demonstrated the ability of reaching ultra high mechanical frequencies in NEMS resonators by combining the beneficial mechanical properties of diamond with unique resonator geometry. In addition, the quality factors of such resonators are not compromised due to localized strain and enhanced through mechanical impedance mismatch. The polycrystalline nature of the diamond films is observed to introduce significant scatter in the resulting stiffness and frequency characteristics. In addition to the reported high frequency and quality factors, these multi-mode resonators can be used to explore other mechanical phenomena such as non-linear mode coupling, which is the subject of ongoing research.

Dr. Xingcheng Xiao's assistance with the development of the seeding and growth procedures is gratefully acknowledged. Diamond fabrication at Brown University is supported by NSF (DMR-0305418). The work at Boston University is supported by NSF (DMR-0449670 and CCF-0432089).

[a] *E-mail: mohanty@buphy.bu.edu*